\magnification=1200
\baselineskip=13pt
\overfullrule=0pt
\tolerance=100000
%%%%%%%%%%%%%%%%%%%%%%%%%%%%%%%%%%%%%%%%%%%%%%%%%%%%%%%%%%%%%%%%%%%%%%%%%%%%%%%

\font\tenbifull=cmmib10 \skewchar\tenbifull='177
\font\tenbimed=cmmib7   \skewchar\tenbimed='177
\font\tenbismall=cmmib5  \skewchar\tenbismall='177
\textfont9=\tenbifull
\scriptfont9=\tenbimed
\scriptscriptfont9=\tenbismall

\mathchardef\alpha="710B
\mathchardef\beta="710C
\mathchardef\gamma="710D
\mathchardef\delta="710E
\mathchardef\epsilon="710F
\mathchardef\zeta="7110
\mathchardef\eta="7111
\mathchardef\theta="7112
\mathchardef\iota="7113
\mathchardef\kappa="7114
\mathchardef\lambda="7115
\mathchardef\mu="7116
\mathchardef\nu="7117
\mathchardef\micron="716F
\mathchardef\xi="7118
\mathchardef\pi="7119
\mathchardef\rho="711A
\mathchardef\sigma="711B
\mathchardef\tau="711C
\mathchardef\upsilon="711D
\mathchardef\phi="711E
\mathchardef\chi="711F
\mathchardef\psi="7120
\mathchardef\omega="7121
\mathchardef\varepsilon="7122
\mathchardef\vartheta="7123
\mathchardef\varphi="7124
\mathchardef\varrho="7125
\mathchardef\varsigma="7126
\mathchardef\varpi="7127
%%%%%%%%%%%%%%%%%%%%%%%%%%%%%%%%%%%%%%%%%%%%%%%%%%%%%%%%%%%%%%%%%%%%%%%%%%%%%

\font\pq=cmr10 at 8truept

\
\def\value{\Big|_{\matrix{\scriptscriptstyle J_0=(D\Phi_0)\cr
\noalign{\vskip -3pt}%
\scriptscriptstyle J_1=(D\Phi_1)}}}
{\hfill \hbox{\vbox{\settabs 1\columns
\+ hep-th/9704126\cr
}}}
\centerline{}
\bigskip
\bigskip
\bigskip
\baselineskip=18pt

\centerline{\bf The sTB-B  Hierarchy}
\vfill
{\baselineskip=11pt
\centerline{J. C. Brunelli\footnote{*}{e-mail address: brunelli@fsc.ufsc.br}}
\medskip
\medskip
\centerline{Universidade Federal de Santa Catarina}
\centerline{Departamento de F\'\i sica -- CFM}
\centerline{Campus Universit\'ario -- Trindade}
\centerline{C.P. 476, CEP 88040-900}
\centerline{Florian\'opolis, SC -- BRAZIL}
\medskip
\medskip
\centerline{and}
\medskip
\medskip
\centerline{Ashok Das}
\medskip
\medskip
\centerline{Department of Physics and Astronomy}
\centerline{University of Rochester}
\centerline{Rochester, NY 14627 -- USA}
}
\vfill

\centerline{\bf {Abstract}}

\medskip
\medskip
We construct a new supersymmetric two boson (sTB-B) hierarchy and study its
properties. We derive the conserved quantities and the Hamiltonian structures
(proving the Jacobi identity) for the system. We show how this system gives the sKdV-B
equation and its Hamiltonian structures upon appropriate reduction. We also
describe the zero curvature formulation of this hierarchy both in the
superspace as well as in components.
 
\vfill
\eject
\eject
\bigskip
\noindent {\bf 1. {Introduction}}
\medskip

In the last few years integrable models [1-3] have become increasingly 
relevant in the study of strings through the matrix models. It was shown 
that the KdV hierarchy appears in the so called double scaling limit of the 
one-matrix model. This fact allowed the exact computation of correlation 
functions on arbitrary topology [4]. However, since supersymmetric string 
theories are believed to be more fundamental, a generalization of these 
techniques to the supersymmetric case is in order.
 A first step in this direction was pursued in [5] and an arbitrary genus 
solution was found in [6]. In fact, a new supersymmetric extension of the KdV 
hierarchy was obtained in the double scaling limit. The authors of Ref. [7,8] 
showed that this new supersymmetric hierarchy, the sKdV-B hierarchy, was a 
sort of supersymmetric covariantization of the bosonic KdV hierarchy. In fact, 
their observation provides us  with a general scheme of supersymmetrization 
for any bosonic hierarchy.

In a parallel study, much attention was devoted to the supersymmetric extension
of the Two-Boson system (TB) [9-11]. The supersymmetric Two-Boson 
(sTB) hierarchy [12,13], has a rich structure. One of its more important properties is 
that it gives rise
 to many other supersymmetric integrable systems under appropriate reduction. 
Our aim in this letter is to supersymmetrize the TB system following the 
procedure  in Refs. [7] and [8]. In this way we will obtain a new 
supersymmetric TB hierarchy, the sTB-B hierarchy and study its properties. 
Under appropriate reduction we will show that the sKdV-B equation results from this system.

The paper is organized as follows. In Sec. 2 we review briefly the Two-Boson 
system. In Sec. 3 we construct the new supersymmetric Two-Boson system  showing
its bi-Hamiltonian structure. We also explain how to prove the Jacobi identity 
for odd Poisson structures 
using super prolongation techniques.  In Sec. 4 we show how the sKdV-B equation
is embedded in this system. We also show that a supersymmetric generalization 
of the nonlinear Schr\"odinger equation (NLS) using this scheme fails to give 
a local equation.
In Sec. 5 we construct the zero curvature formulation for the sTB-B hierarchy 
in the superspace as well as in components. Our conclusions are presented in 
Sec. 6. We also refer the readers to [13] where details about supersymmetric 
calculations can be found.
%vfill\eject

\bigskip
\noindent {\bf 2. {The TB Hierarchy}}
\medskip

The two boson system is given by the equations [9-11] (prime denotes derivative
with respect to $x$)
$$
\eqalign{
{\partial  J_0 \over \partial  t} &= (2 J_1 + J_0^2 -
J^\prime_0 )^\prime\cr
\noalign{\vskip 4pt}%
{\partial  J_1 \over \partial  t} &= (2 J_0 J_1 +
J^\prime_1 )^\prime\cr
}\eqno(1)
$$
This system has a nonstandard Lax representation [9,14]
$$
{\partial  L \over \partial  t} = \left[ L, \left( L^2 \right)_{\geq 1}
\right] \eqno(2)
$$
with a Lax operator given by
$$
L =  \partial  - J_0 + \partial^{-1} J_1 \eqno(3)
$$
In fact, equations (1) are part of an hierarchy of equations which can be expressed in bi-Hamiltonian form 
$$
\pmatrix{{\partial J_0\over\partial t_k}\cr
\noalign{\vskip 10pt}%
{\partial J_1\over\partial t_k}}={\cal D}_1
\pmatrix{{\delta H_{k+1}\over\delta J_0}\cr
\noalign{\vskip 10pt}%
{\delta H_{k+1}\over\delta J_1}}=
{\cal D}_2
\pmatrix{{\delta H_{k}\over\delta J_0}\cr
\noalign{\vskip 10pt}%
{\delta H_{k}\over\delta J_1}}
\eqno(4)
$$

The conserved charges (Hamiltonians) are given by
$$
H_n = \ {\rm Tr}\ L^n = \int dx\ {\rm Res}\ L^n =\int dx\,h_n(J_0,J_1)\qquad \qquad
n = 1,2,3,\dots \eqno(5)
$$
where \lq\lq Res" stands for the coefficient of the $\partial^{-1}$ term in the
pseudo-differential operator. The first few conserved charges are
$$
\eqalign{
H_1 &= \int dx\ J_1\cr
H_2 &= \int dx\ J_0 J_1\cr
H_3 &= \int dx\ \left( J^2_1 - J_0^\prime J_1 + J_1 J^2_0 \right)\cr
}\eqno(6)
$$
and the two  Hamiltonian structures for the system in eq. (4) are given by
$$
\eqalign{
{\cal D}_1=& \pmatrix{0 &\partial\cr
\noalign{\vskip 5pt}%
\partial&0}\cr
{\cal D}_2=& \pmatrix{2\partial &\partial J_0 - \partial^2\cr
\noalign{\vskip 12pt}%
J_0\partial+\partial^2&J_1\partial+\partial J_1}\cr
}\eqno(7)
$$
For $k=2$ we obtain the equations (1) from (4).

\bigskip
\noindent {\bf 3. {The sTB-B Hierarchy and Its Integrability}}
\medskip

The standard procedure to obtain the new supersymmetric TB hierarchy 
(hereafter referred to as the sTB-B hierarchy) is to first replace the 
variables $J_0$, $J_1$ in the TB hierarchy (4) (which we will denote by 
TB($J_0$,$J_1$) following Ref. [7]) by the superfields
$$
\eqalign{(D\Phi_0) &= J_0 + \theta \psi_0'\cr
(D\Phi_1) &= J_1 + \theta \psi_1' \cr
}
\eqno(8)
$$
In this way we get a new hierarchy which we will denote by 
TB($(D\Phi_0)$,$(D\Phi_1)$) and the new sTB-B($\Phi_0$,$\Phi_1$) hierarchy is 
obtained by taking off one derivative from the left and the right sides of the 
equation. So, from (4) the TB($(D\Phi_0)$,$(D\Phi_1)$) hierarchy is
$$
\pmatrix{{\partial (D\Phi_0)\over\partial t_k}\cr
\noalign{\vskip 10pt}%
{\partial (D\Phi_1)\over\partial t_k}}={\cal D}_1\value
\pmatrix{{\delta H_{k+1}\over\delta J_0}\value\cr
\noalign{\vskip 10pt}%
{\delta H_{k+1}\over\delta J_1}\value}=
{\cal D}_2\value
\pmatrix{{\delta H_{k}\over\delta J_0}\value\cr
\noalign{\vskip 10pt}%
{\delta H_{k}\over\delta J_1}\value}
\eqno(9)
$$
Using, for instance, the result from Lemma 10 in Ref. [8] we have
$$
\eqalign{
{\delta\ \over\delta\Phi_0}\int dx d\theta\, h_n((D\Phi_0),(D\Phi_1))=&
D{\delta\ \over\delta J_0}\int dx\,h_n(J_0,J_1)\value\cr
{\delta\ \over\delta\Phi_1}\int dx d\theta\, h_n((D\Phi_0),(D\Phi_1))=&
D{\delta\ \over\delta J_1}\int dx\,h_n(J_0,J_1)\value\cr
}\eqno(10)
$$
and (9) yields
$$
\pmatrix{{\partial \Phi_0\over\partial t_k}\cr
\noalign{\vskip 10pt}%
{\partial \Phi_1\over\partial t_k}}={\cal J}_1
\pmatrix{{\delta K_{k+1}\over\delta \Phi_0}\cr
\noalign{\vskip 10pt}%
{\delta K_{k+1}\over\delta \Phi_1}}=
{\cal J}_2
\pmatrix{{\delta K_{k}\over\delta \Phi_0}\cr
\noalign{\vskip 10pt}%
{\delta K_{k}\over\delta \Phi_1}}
\eqno(11)
$$
where
$$
\eqalign{
{\cal J}_1=& D^{-1}{\cal D}_1\value D^{-1}=\pmatrix{0 & 1\cr
\noalign{\vskip 5pt}%
1 & 0}\cr
{\cal J}_2=&D^{-1}{\cal D}_2\value D^{-1}=\pmatrix{2&
D(D\Phi_0)D^{-1}-D^2\cr
\noalign{\vskip 20pt}%
D^{-1}(D\Phi_0)D+D^2&D^{-1}(D\Phi_1)D+D(D\Phi_1)D^{-1}}
}\eqno(12)
$$
and
$$
K_n=\int dx d\theta\, h_n((D\Phi_0),(D\Phi_1))\eqno(13)
$$

Now, ${\cal J}_1$ and ${\cal J}_2$ are odd Poisson structures and this is a 
characteristic feature of this supersymmetrization scheme.  ${\cal J}_1$ and 
${\cal J}_2$ are symmetric and  ${\cal J}_1$ satisfies the Jacobi identity 
trivially. To show that  ${\cal J}_2$ satisfies the Jacobi identity is 
slightly involved and we use the super-prolongation techniques [15,16,13] 
adapted for odd Poisson structures. The modification needed to treat odd 
Poisson structures is to use a two-component column matrix of fermionic 
superfields
$$
{\vec\Omega}=\pmatrix{
\Omega_0\cr
\noalign{\vskip 10pt}%
\Omega_1}\eqno(14)
$$
and construct the bivector associated with the Hamiltonian structure ${\cal J}$
$$
\Theta_{\cal J}={1\over 2}\sum_{\alpha,\beta}\int dz\,
\left(({\cal J})_{\alpha\beta}\Omega_\beta\right)
\wedge\Omega_\alpha\qquad
\alpha,\beta=0,1\eqno(15)
$$
The necessary and sufficient condition for $\cal J$ to define a Hamiltonian structure is that the prolongation of this bivector should vanish
$$
\hbox{\bf pr}\,{\vec v}_{{\cal J}{\vec\Omega}}
(\Theta_{{\cal J}}) = 0\eqno(16)
$$
For ${\cal J}_2$ given by (12) we get
$$
\eqalign{
\hbox{\bf pr}\,{\vec v}_{{\cal J}_2{\vec\Omega}}(\Phi_0)=&
2\Omega_0+(D^2\Phi_0)(D^{-1}\Omega_1)+(D\Phi_0)\Omega_1-(D^2\Omega_1)\cr
\noalign{\vskip 10pt}%
\hbox{\bf pr}\,{\vec v}_{{\cal J}_2{\vec\Omega}}(\Phi_1)=&\ 
\left(D^{-1}(D\Phi_0)(D\Omega_0)\right)+(D^2\Omega_0)+
\left(D^{-1}(D\Phi_1)(D\Omega_1)\right)\cr
&+(D^2\Phi_1)(D^{-1}\Omega_1)+(D\Phi_1)\Omega_1\cr
}\eqno(17)
$$
and using this it is  easy to show that the prolongation of the bivector (15) vanishes
$$
\hbox{\bf pr}\,{\vec v}_{{\cal J}_2{\vec\Omega}}
(\Theta_{{\cal J}_2}) = 0\eqno(18)
$$
leading to the fact  that ${\cal J}_2$ satisfies the Jacobi identity. Also, 
it can be shown that ${\cal J}_1$ and ${\cal J}_2$ are compatible since
$$
\hbox{\bf pr}\,{\vec v}_{({\cal J}_2+\alpha{\cal J}_1){\vec\Omega}}
(\Theta_{{\cal J}_2+\alpha{\cal J}_1}) = 0\eqno(19)
$$
In this way the sTB-B hierarchy (11) is a bi-Hamiltonian system, which implies its integrability according to Magri's theorem [15].

From the first Hamiltonian structure in (11) we can see that all flows are local. For $k=2$ ($t_2=t$) we get the equations
$$
\eqalign{{\partial  \Phi_0 \over \partial  t} &=
 - (D^4 \Phi_0) + (D(D\Phi_0)^2)+ 2(D^2 \Phi_1)\cr
\noalign{\vskip 4pt}%
{\partial  \Phi_1 \over \partial  t} &=
 (D^4 \Phi_1) + 2(D((D\Phi_0)(D\Phi_1)))\cr
}\eqno(20)
$$
which we call the sTB-B equation. Note that these equations are different from the sTB [12,13] equation only in the last term in the equation for $\Phi_1$. In components (20) yields
$$
\eqalign{
{\partial  J_0 \over \partial  t} &= ( 2J_1 + J_0^2 - J_0^\prime)^\prime\cr
\noalign{\vskip 4pt}%
{\partial  \psi_0 \over \partial  t} &= 2 \psi_1^\prime + 2 \psi_0^\prime
J_0 - \psi_0^{\prime \prime}\cr
\noalign{\vskip 4pt}%
{\partial  J_1 \over \partial  t} &= ( 2J_0 J_1 + J_1^\prime)^\prime\cr
\noalign{\vskip 4pt}%
{\partial  \psi_1 \over \partial  t} &=\psi_1^{\prime\prime}+ 2 \psi_1^\prime J_0 + 2J_1\psi_0^\prime\cr
}
\eqno(21)
$$
which, of course, is invariant under the supersymmetry transformations
$$
\eqalign{
\delta  J_0 &= \epsilon \psi_0^\prime\cr
\delta  J_1 &= \epsilon \psi_1^\prime\cr
\delta \psi_0 &= \epsilon J_0\cr
\delta  \psi_1 &= \epsilon J_1\cr
}
\eqno(22)
$$

Equation (20) can also be obtained from the Lax operator
$$
L = D^2 -(D \Phi_0) + D^{-2} (D\Phi_1)
\eqno(23)
$$
with the nonstandard Lax representation (2). The conserved charges (13) can be written as
$$
K_n=\int dz\,\hbox{Res}\,L^n\qquad n=1,2,3,\dots \eqno(24)
$$
where ``Res'' stands for the residue which is defined to be the
coefficient of the $D^{-2}=\partial^{-1}$ term in the pseudo super-differential
operator. These charges are purely fermionic. The first ones are
$$
\eqalign{
K_1=&\int dz\, (D\Phi_1)\cr
K_2=&\int dz\, (D\Phi_0)(D\Phi_1)\cr
K_3=&\int dz\,\Bigl[(D\Phi_1)^2-(D^3\Phi_0)(D\Phi_1)+(D\Phi_1)(D\Phi_0)^2\Bigr]\cr
}\eqno(25)
$$
and when written in components they assume the form
$$
\eqalign{
K_1=&\int dx\, \psi_1'=0\cr
K_2=&\int dx\, \Bigl(\psi_0'J_1+J_0\psi_1'\Bigr)\cr
K_3=&\int dx\,\Bigl(2J_1\psi_1'-\psi_0''J_1-J_0'\psi_1'+\psi_1'J_0^2+2J_1J_0\psi_0'\Bigr)\cr
}\eqno(26)
$$
which can be easily checked to be invariant under supersymmetry transformations
(22). Note that, in the bosonic limit these charges vanish. 

Let us also note that the charges  $K_n$ can be simply written as the
supersymmetric variation of $H_{n}$ of (6) (without $\epsilon$), namely,
$$
K_n=\delta H_n\eqno(27)
$$
In fact, this allows us to identify the bosonic TB conserved charges as the 
$\theta$ independent part of $\hbox{Tr}\,L^n$. We could also have constructed 
bosonic conserved charges for the sTB-B system using odd powers of the square 
root of (23) [17,18]
$$
Q_{2n-1\over2}=\int dz\,\hbox{Res}\,L^{2n-1\over 2}\qquad n=1,2,3,\dots \eqno(28)
$$
However, as pointed out by Manin and Radul [19],  $L^{1/2}$, in such a case, is
nonunique in general. On the other hand, if we require the coefficient
functions to vanish at spatial infinity, these bosonic charges are unique and
we have checked that they are conserved as well. These are, in fact, a new set
of charges distinct from $H_{n}$ [20]. 

\eject
\bigskip
\noindent {\bf 4. {Reductions of the sTB-B Hierarchy}}
\medskip

The TB  as well the sTB systems reduce to various known integrable systems. 
Let us show that we can embed the sKdV-B equation into the sTB-B system. 
Following Refs. [13,21] and using (23), we obtain
$$
(L^3)_{\geq 1} = D^6 + 3( D \Phi_1 )D^2 -3(D^3\Phi_0)D^2-3(D\Phi_0)D^4
+3(D\Phi_0)^2D^2
\eqno(29)
$$
The nonstandard Lax equation
$$
{\partial   L \over \partial  t} = [ L, (L^3)_{\geq 1}]
\eqno(30)
$$
leads to the equations
$$
\eqalign{
{\partial\Phi_1\over\partial  t} =&
-(D^6\Phi_1 ) - 3 D \biggl((D \Phi_1)^2+(D\Phi_1)(D\Phi_0)^2+
(D^3\Phi_1)(D\Phi_0)-2(D\Phi_1)(D^3\Phi_0)\biggr)\cr
{\partial\Phi_0\over\partial  t} =&
-(D^6\Phi_0)-D\biggl(6(D\Phi_1 )(D\Phi_0)-3(D\Phi_0)(D^3\Phi_0)+
(D\Phi_0)^3\biggr)\cr
}\eqno(31)
$$
which, for
$$
\eqalign{\Phi_0 & =  0\cr
\Phi_1 & = \Phi} \eqno(32)
$$
gives
$$
{\partial\Phi\over\partial  t} =
-(D^6\Phi) - 3 D\left((D\Phi)^2\right)\eqno(33)
$$
This is, indeed, the sKdV-B equation considered in [7]. The Hamiltonian 
structures of the sKdV-B can also be obtained from the sTB-B ones. Following 
the Dirac reduction [21], let us indicate by $\overline{{\cal O}}$ quantities  
${\cal O}$ constrained to satisfy (32). Then,
$$
{\cal J}^{\,\hbox{\pq sKdV-B}}={\overline{\cal J}}_{22}-
{\overline{\cal J}}_{21}{\overline{\cal J}}_{11}^{-1}{\overline{\cal J}}_{12}
\eqno(34)
$$
>From (12), it follows that
$$
({\overline{\cal J}}_2)^{-1}_{11}={1\over2}\eqno(35)
$$
and, consequently, (34) yields
$$
{\cal J}_2^{\,\hbox{\pq sKdV-B}}={1\over2}
(D^4+2D^{-1}(D\Phi)D +2D(D\Phi)D^{-1})\eqno(36)
$$

As was noted in [21], the first Hamiltonian for the sKdV-B needs to be obtained
from the sTB-B  Hamiltonian structure ${\cal J}_{0}$ and not from ${\cal J}_1$.
Recursively we have
$$
{\cal J}_0=R^{-1}{\cal J}_1\eqno(37)
$$
where $R={\cal J}_2{\cal J}_1^{-1}$ is the recursion operator. It is straightforward to show that
$$
{\overline R}^{-1}=\pmatrix{-{1\over2}\left({\cal J}_2^{\,\hbox{\pq
sKdV-B}}\right)^{-1}D^2&
\left({\cal J}_2^{\,\hbox{\pq sKdV-B}}\right)^{-1}\cr
\noalign{\vskip 20pt}%
{1\over 2}-{1\over 4}D^2\left({\cal J}_2^{\,\hbox{\pq sKdV-B}}\right)^{-1}D^2&
{1\over2}D^2\left({\cal J}_2^{\,\hbox{\pq sKdV-B}}\right)^{-1}
}\eqno(38)
$$
Therefore, it follows from (37) and (34) that
$$
{\cal J}_1^{\,\hbox{\pq sKdV-B}}=({\overline{\cal J}}_0)_{22}-
({\overline{\cal J}}_0)_{21}({\overline{\cal J}}_0)^{-1}_{11}
({\overline{\cal J}}_0)_{12}={1\over 2}\eqno(39)
$$
The Hamiltonian structures (36) and (39) are, indeed, the ones derived in [7].

It is well known [10,11,13] that the field redefinitions
$$
\eqalign{
J_0=&-{q'\over q}=-(\ln q )'\cr
J_1=&{\bar q}q
}\eqno(40)
$$
take the TB equation (1) to the nonlinear Schr\"odinger equation (NLS)
$$
\eqalign{
{\partial q\over\partial t}=&-(q''+2({\bar q}q)q)\cr
\noalign{\vskip 4pt}%
{\partial {\bar q}\over\partial t}=&\,\,{\bar q}''+2({\bar q}q){\bar q}\cr
}\eqno(41)
$$
So, it will be expected that some field redefinition will take the sTB-B equation to a new system, the sNLS-B equation. Let us use the field redefinition
$$
\eqalign{
(D\Phi_0)=&-{(D^3Q)\over (DQ)}\cr
(D\Phi_1)=&(DQ)(D{\overline Q})
}\eqno(42)
$$
where
$$
\eqalign{(DQ )&= q+ \theta \phi'\cr
{(D\overline Q}) &= {\bar q} + \theta {\overline \phi}' \cr
}
\eqno(43)
$$
Using this in the Lax operator (23), we obtain
$$
L=D^2+{(D^3Q)\over(DQ)}+D^{-2}(DQ)(D{\overline Q})=G{\widetilde L}G^{-1}
\eqno(44)
$$
where
$$
\eqalign{
G=&(DQ)^{-1}\cr
{\widetilde L}=&D^2+(DQ)D^{-2}(D{\overline Q}) \cr
}\eqno(45)
$$
So, we have the Lax operators $L$ and $\widetilde L$ related by a gauge transformation. Taking the formal adjoint of  $\widetilde L$
$$
{\cal L}={\widetilde L}^*=-D^2+(D{\overline Q})D^{-2}(DQ)\eqno(46)
$$  
we can obtain consistent equations using the standard Lax representation
$$
{\partial{\cal L}\over\partial t}=\left[
\left({\cal L}^2\right)_{+},{\cal L}\right]
\eqno(47)
$$
and they are
$$
\eqalign{
{\partial  Q \over \partial  t} &=-(D^4 Q) + 2D^{-1}((D{\overline Q})(DQ)^2)\cr
\noalign{\vskip 4pt}%
{\partial  \overline Q \over \partial  t} &= (D^4 \overline Q )-
2D^{-1}((D{\overline Q})^2(DQ))\cr
}\eqno(48)
$$
which, unfortunately is nonlocal. The reader can verify that the same result (48) can be obtained directly from the supersymmetrization of the bosonic NLS hierarchy following the same steps from (8) to (11) used for the TB hierarchy.

\bigskip
\noindent {\bf 5. {Zero Curvature}}
\medskip

As is clear from the discussions of the earlier sections, the Hamiltonian
structures of the sTB-B are fermionic. Normally, the Hamiltonian structures of
the integrable models correspond to interesting symmetry algebras and dictate
the choice of the gauge group in the zero curvature formulation of the problem 
[22]. It is,
therefore, interesting to ask whether a zero curvature formulation of this
model can be achieved. To this end, we first discuss the problem in superspace
before returning to the component formulation.

From the structure of the Lax operator in  (23), we note that the linear
problem associated with the sTB-B system is given by ($\lambda$ is the constant
spectral parameter.)
$$
L\chi_{1} =\lambda \chi_{1}\eqno(49)
$$
where $\chi_{1}$ is assumed to be a bosonic superfield eigenfunction. We note
that we can write the equation in a completely local form by introducing a
second superfield as
$$
\chi_{2} = [D^{-2} (D\Phi_{1}) \chi_{1}]\eqno(50)
$$
so that the linear equation can be written as
$$
\partial_{x} \chi = {\cal A}_{1} \chi\eqno(51)
$$
where
$$
\eqalign{
\chi & = \pmatrix{\chi_{1} \cr\chi_{2}}\cr
{\cal A}_{1} & = \pmatrix{\lambda + (D\Phi_{0}) & -1\cr
(D\Phi_{1}) & 0}
}\eqno(52)
$$
It is interesting to note here that both the superfields $\chi_{1}$ and
$\chi_{2}$ are bosonic. This is different from the discussion in [23] and it is
not clear how a graded symmetry group would arise in such a case.

Writing the time evolution equation as
$$
\partial_{t} \chi = {\cal A}_{0} \chi\eqno(53)
$$
where we assume that, in terms of general superfields, ${\cal A}_{0}$ has the
form
$$
{\cal A}_{0} = \pmatrix{A & B \cr C & E}\eqno(54)
$$
the zero curvature equation
$$
\partial_{t} {\cal A}_{1} - \partial_{x} {\cal A}_{0} - [{\cal A}_{0}, {\cal
A}_{1}] = 0\eqno(55)
$$
is obtained as the compatibility condition between the two. This leads to the
constraint conditions
$$
\eqalign{
A & = E + B_{x} - (\lambda + (D\Phi_{0}))B \cr
C & = E_{x} - B (D\Phi_{1})}\eqno(56)
$$
as well as the two dynamical equations
$$
\eqalign{
{\partial\ \over\partial t} (D\Phi_{0})& = B_{xx} + 2 E_{x} -
[B(D\Phi_{0})]_{x} - \lambda B_{x} \cr
 {\partial\ \over\partial t}(D\Phi_{1})& =  E_{xx} + E_{x} (D\Phi_{0}) - 2
B_{x} (D\Phi_{1}) - B (D^{3}\Phi_{1}) + \lambda E_{x}}
\eqno(57)
$$
It is clear now that if we write a Taylor series expansion of the forms
$$
B = \sum_{n=0}^{N} \lambda^{N-n} B_{n}\,,{\hskip .3in} E = \sum_{n=0}^{N}
\lambda^{N-n} E_{n}\eqno(58)
$$
the recursion relations take the form
$$
\pmatrix{(DB_{n+1})\cr (DE_{n+1})}=\pmatrix{
D^{2} - D (D\Phi_{0}) D^{-1} & 2 \cr
D (D\Phi_{1}) D^{-1} + D^{-1} (D\Phi_{1}) D & - D^{2} - D^{-1} (D\Phi_{0}) D\cr} \pmatrix{(DB_{n})\cr(DE_{n})}\eqno(59)
$$
which is the same as the recursion operator following from  Eqs. (12) with appropriate identifications. If
we now choose
$$
B_{0} = 1\,, {\hskip .5in} E_{0} = 0\eqno(60)
$$
it is straightforward to check that the sTB-B hierarchy follows. In particular,
the third order equation yields the sKdV-B equation when $\Phi_{0}=0$ as it
should.

From the form of ${\cal A}_{1}$, the symmetry group appears more like
$SL(2)\times U(1)$. A graded symmetry algebra is not at all clear and the
reduction of the zero curvature condition to components is even more obscure.
We have, however, succeed in obtaining the zero curvature condition for the
hierarchy in terms of $4\times 4$ matrices. Without going into details, we
simply note that the matrices
$$
\eqalign{
{\cal A}_{1} & =\pmatrix{
(\lambda + J_{0}) & -1 & 0 & 0 \cr
J_{1} & 0 & 0 & 0 \cr
\psi_{0}' & 0 & (\lambda + J_{0}) & -1 \cr
\psi_{1}' & 0 & J_{1} & 0}\cr
{\cal A}_{0} & =\pmatrix{
F & G & 0 & 0 \cr
H & K & 0 & 0 \cr
P & Q & F & G \cr
R & S & H & K}}
\eqno(61)
$$
would yield, from the zero curvature condition, the constraints (both ${}'$ and
the subscript $x$ denote derivative with respect to $x$)
$$
\eqalign{
F & =  K + G_{x} - (\lambda + J_{0}) G\cr
H & =  K_{x} - G J_{1}\cr
P & =  S + Q_{x} - (\lambda + J_{0}) Q - \psi_{0}' G\cr
R & =  S_{x} - J_{1} Q - \psi_{1}' G}
\eqno(62)
$$
as well as the dynamical equations
$$
\eqalign{
J_{0 t} & =  G_{xx} + 2 K_{x} - J_{0 x} G - (\lambda + J_{0}) G_{
x}\cr
J_{1 t} & = K_{xx} - J_{1 x} G - 2J_{1} G_{x} + (\lambda + J_{0})
K_{x}\cr
\psi_{0 t}' & =  Q_{xx} + 2 S_{x} - J_{0 x} Q - (\psi_{0}' G)' - (\lambda +
J_{0}) Q_{x}\cr
\psi_{1 t}' & = S_{xx} - J_{1 x} Q - 2 J_{1} Q_{x} - 2 \psi_{1}' G_{x} -
\psi_{1}'' G + \psi_{0}' K_{x} + (\lambda + J_{0}) S_{x}}
\eqno(63)
$$
Once again, writing a Taylor series as in Eq. (58) and choosing
$$
G_{0} = 1\,,{\hskip .5in} K_{0} = 0 = Q_{0} = S_{0}\eqno(64)
$$
leads to the sTB-B hierarchy in components. It is worth noting here that the
matrix structure, in components, suggests an underlying symmetry algebra of
$OSp(2|2)$ which is also isomorphic to $SL(2|1)$.
\bigskip
\noindent {\bf 6. {Conclusions}}
\medskip
We have constructed a new supersymmetric two boson (sTB-B) hierarchy and
studied its properties. We derived the conserved charges and have shown that it
is a bi-Hamiltonian system. The Jacobi identity for the Hamiltonian structures
are verified using the super-prolongation technique. We have shown how this
system gives the sKdV-B system as well as its Hamiltonian structures upon 
appropriate reduction. We have also described the zero curvature formulation of
this new hierarchy both in the super space as well as in components and have
brought out its distinguishing features.
\bigskip
\leftline{\bf Acknowledgments}
\medskip

J.C.B. was supported by CNPq, Brazil. A.D. was supported in part by the U.S. 
Department of Energy Grant No. DE-FG-02-91ER40685 and by NSF-INT-9602559. 

\vfill\eject

\leftline{\bf References}
\bigskip

\item{1.} L.D. Faddeev and L.A. Takhtajan, ``Hamiltonian Methods in
the Theory of Solitons'' (Springer, Berlin, 1987).

\item{2.} A. Das, ``Integrable Models'' (World Scientific, Singapore,
1989).

\item{3.} L. A. Dickey, ``Soliton Equations and Hamiltonian Systems'' (World
Scientific, Singapore, 1991).

\item{4.} D. J. Gross and A. A. Midgal, Phys. Rev. Lett. {\bf 64}, 127 (1990);
D. J. Gross and A. A. Midgal, Nucl. Phys. {\bf B340}, 333 (1990); E. Br\'ezin
and V. A. Kazakov, Phys. Lett. {236B}, 144 (1990); M. Douglas and S. H.
Shenker,
Nucl. Phys. {\bf B335}, 635 (1990);
A. M. Polyakov in ``Fields, Strings and Critical Phenomena'', Les
Houches 1988, ed. E. Br\'ezin and J. Zinn-Justin (North-Holland, Amsterdam,
1989); L. Alvarez-Gaum\'e, Helv. Phys. Acta {\bf 64}, 361 (1991); P. Ginsparg
and G. Moore, ``Lectures on 2D String Theory and 2D Gravity'' (Cambridge, New
York, 1993).

\item{5.} L. Alvarez-Gaum\'e and J. L. Man\~es, Mod. Phys. Lett. {\bf A6},
2039 (1991); L. Alvarez-Gaum\'e, H. Itoyama, J. Man\~es and A. Zadra, Int. J.
Mod. Phys. {\bf A7}, 5337 (1992).

\item{6.}  K. Becker and M. Becker, Mod. Phys. Lett. {\bf A8}, 1205, (1993).

\item{7.}  J. M. Figueroa-O'Farrill and S. Stanciu, Phys. Lett. {\bf B316}, 282 (1993). 

\item{8.}  J. M. Figueroa-O'Farrill and S. Stanciu, Mod. Phys. Lett. {\bf A8}, 2125 (1993). 

\item{9.} B.A. Kupershmidt, Commun. Math. Phys. {\bf 99}, 51 (1985).

\item{10.} H. Aratyn, L.A. Ferreira, J.F. Gomes and A.H. Zimerman, Nucl.
Phys. {\bf B402}, 85 (1993); H. Aratyn, L.A. Ferreira, J.F. Gomes and A.H.
Zimerman, ``On $W_\infty$ Algebras, Gauge Equivalence of KP Hierarchies,
Two-Boson Realizations and their KdV Reductions'', in Lectures at the VII
J. A. Swieca Summer School, S\~ao Paulo, Brazil, January 1993,
eds. O. J. P. \'Eboli and V. O. Rivelles (World Scientific, Singapore, 1994);
H. Aratyn, E. Nissimov and S. Pacheva, Phys. Lett. {\bf B314}, 41 (1993).

\item{11.} L. Bonora and C.S. Xiong, Phys. Lett. {\bf B285}, 191 (1992); L.
Bonora and C.S. Xiong, Int. J. Mod. Phys. {\bf A8}, 2973 (1993).

\item{12.} J. C. Brunelli and A. Das, Phys. Lett. {\bf B337}, 303 (1994).

\item{13.} J. C. Brunelli and A. Das, Int. J. Mod. Phys. {\bf A10}, 4563 (1995).

\item{14.} J. C. Brunelli, A. Das and W.-J. Huang, Mod. Phys. Lett. {\bf 9A},
2147 (1994).

\item{15.} P. J. Olver , ``Applications of Lie Groups to Differential
Equations'', Graduate Texts in Mathematics, Vol. 107 (Springer, New York,
1986).

\item{16.} P. Mathieu, Lett. Math. Phys. {\bf 16}, 199 (1988).

\item{17.} P. Dargis and P. Mathieu, Phys. Lett. {\bf A176}, 67 (1993).

\item{18.} J. C. Brunelli and A. Das, Phys. Lett. {\bf B354}, 307 (1995).

\item{19.} Y. I. Manin and A. O. Radul, Commun. Math. Phys. {\bf 98}, 65 (1985).

\item{20.} J. C. Brunelli and A. Das, work in progress.

\item{21.} J. C. Brunelli and A. Das, Mod. Phys. Lett. {\bf A10}, 2019 (1995).

\item{22.} A. Das and S. Roy, J. Math. Phys. {\bf 31}, 2145 (1990); A. Das, W.
-J. Huang and S. Roy, Int. J. Mod. Phys. {\bf A7}, 3447 (1992); A. Das and S.
Roy, Mod. Phys. Lett. {\bf A11}, 1317 (1996).

\item{23.} H. Aratyn, A. Das and C. Rasinariu, \lq\lq Zero Curvature Formalism
for Supersymmetric Integrable Hierarchies in Superspace",  UICHEP-TH-97-5.

\end

\item{\ .} Y. I. Manin and A. O. Radul, Commun. Math. Phys. {\bf 98}, 65 (1985).

\item{\ .} J. C. Brunelli and A. Das,  Rev. Math. Phys. {\bf 7}, 1181 (1995).

\item{\ .} J. C. Brunelli and A. Das, Phys. Lett. {\bf B354}, 307 (1995).

\item{\ .} M. Freeman and P. West, Phys. Lett. {\bf 295B}, 59 (1992).

\item{\ .} J. Schiff, ``The Nonlinear Schr\"odinger Equation and
Conserved Quantities in the Deformed Parafermion and SL(2,{\bf R})/U(1)
Coset Models'', Princeton preprint IASSNS-HEP-92/57 (1992)
(also hep-th/9210029).

\item{\ .} A. Das and W.-J. Huang, J. Math. Phys. {\bf 33}, 2487 (1992).

\item{\ .} P. Mathieu, J. Math. Phys. {\bf 29}, 2499 (1988).

\item{\ .} J. M. Figueroa-O'Farrill, J. Mas and E. Ramos, Rev. Math. Phys.
{\bf 3}, 479 (1991).

\item{\ .} J. Barcelos-Neto and A. Das, J. Math. Phys. {\bf 33}, 2743 (1992).

\item{\ .} G.H.M. Roelofs and P.H.M. Kersten, J. Math. Phys. {\bf 33}, 2185
(1992).

\item{\ .} J. C. Brunelli and A. Das, J. Math. Phys. {\bf 36}, 268 (1995).

\item{\ .} P. H. M. Kersten, Phys. Lett. {\bf A134}, 25 (1988).

\item{\ .} N. J. Mackay, Phys. Lett. {\bf B281}, 90 (1992); erratum-ibid.
{\bf B308}, 444 (1993).

\item{ .} E. Abdalla, M. C. B. Abdalla, J. C. Brunelli and A. Zadra, Commun.
Math. Phys. {\bf 166}, 379 (1994).